# A Novel Modified Apriori Approach for Web Document Clustering

Rajendra Kumar Roul[1], Saransh Varshneya[2], Ashu Kalra[2], Sanjay Kumar Sahay[3]

[1,2,3]BITS Pilani K. K. Birla Goa Campus, Zuarinagar, Goa-403726, India

{[1]rkroul,[3]ssahay}@goa.bits-pilani.ac.in
[2]{saranshvarshneya, ashukalra7.17.27}@gmail.com

**Abstract.** The Traditional apriori algorithm can be used for clustering the web documents based on the association technique of data mining. But this algorithm has several limitations due to repeated database scans and its weak association rule analysis. In modern world of large databases, efficiency of traditional apriori algorithm would reduce manifolds. In this paper, we proposed a new modified apriori approach by cutting down the repeated database scans and improving association analysis of traditional apriori algorithm to cluster the web documents. Further we improve those clusters by applying Fuzzy C-Means (FCM), K-Means and Vector Space Model (VSM) techniques separately. For experimental purpose, we use Classic3 and Classic4 datasets of Cornell University having more than 10,000 documents and run both traditional apriori and our modified apriori approach on it. Experimental results show that our approach outperforms the traditional apriori algorithm in terms of database scan and improvement on association of analysis. We found out that FCM is better than K-Means and VSM in terms of F-measure of clusters of different sizes.

## 1 Introduction

World Wide Web is the most important place for Information Retrieval (IR). Today volumes of data on web server are increasing exponentially which is almost doubling in every six months [1]. So we need to have efficient search engines to handle large datasets. The query entered by users to any search engine will bring millions of documents from the web in which many of them are not of user choice. This problem can be further narrow down by making clusters on retrieved documents where each cluster will having similar kind of documents. To identify such similar kinds of documents from a corpus of huge size, association rule of data mining play a vital role. This rule discovers frequent itemset in a large dataset [2]. Association rules states that the knowledge of frequent itemset can be used to find out how an itemset is influenced by the presence of another itemset in the corpus. An itemset is frequent, if it is present in at least $x\%$ of the total transactions $z$ in the database $D$, where $x$ is the support threshold. When the numbers of items included in the database transaction are high and we are finding itemset with small support, the number of frequent itemset found are quite large, and it makes the problem very expensive to solve, both in time and space. Hence the minimum support count affects the computational cost of the higher (say $k^{th}$) iteration of apriori algorithm. Thus, one can say that the cardinality of $C^k$ and the size of $D$ affect the overall computational cost. In fact, the traditional apriori algorithm generated and tested the candidate itemset in a level wise manner [3] using iterative database scan which makes the computational cost very high.

Our approach considered documents as itemset and keywords as transaction so that we end up with clusters with a minimum frequency support threshold. By making keywords as transaction and document as itemset, we combined the support threshold idea of association rule mining algorithm with the output like that of a clustering algorithm. The salient features of our approach in comparison with traditional apriori algorithm can be described as follows:
- only one database scan by bringing the database into memory.
- reduces the time for accessing the transactions.
- completely eliminated the repeated database scan.
- nullifies the transactions which are no longer in use.
- decreasing the number of candidate itemset during the candidate generation step.

We applied this modified apriori technique on a corpus of web documents to get some initial clusters. Next the FCM[2], K-Means[2] and VSM[2] techniques are applied separately to these initial clusters for further strengthening them as it missed out many documents because of the strong association rule of apriori with high minimum support. The experiment has been carried out on Classic3 and Classic4 dataset of Cornell University and the results demonstrate the applicability, efficiency and effectiveness of our approach.

The outline of the paper is organized as follows: Section 2 covers the related work based on different association techniques used for clustering. Background of the proposed work is described in section 3. In section 4, we describe the proposed approach adopted to form the clusters. Section 5 shows the experimental work and finally we concluded the present work with future enhancement in section 6.

## 2  Related Work

Association rule of data mining is used to find out a set of frequent itemset from a list of itemset depending on the minimum support of user choice. It plays a vital role in clustering technique. Wang *et al*.[4] proposed a technique to find out frequent itemset in a large database and to mine association rules from frequent itemset. Though they improved the mining association rule algorithm based on support and confidence but creates useless rules and lost useful rules out of which the useless rules are discarded to discovered more reasonable association rules. Bodon[5] in his work, tried to store data which also stores frequent itemset with candidate itemset. Xiaojun[6] includes grain bit binary extraction to mine frequent itemsets . Wang *et*

*al.*[7] in their approach tried to use normative database of users to get itemset. Xiaohong et al.[8] have used a cross-linker structure instead to substitute array based representation of transaction database. Iva *et al*. [9] applied properties of fuzzy confirmation measures in association rule mining process. Yanping *et al*.[10] uses a two-layer web clustering approach to cluster for a number of web access patterns from web logs. At the first layer, Learning Vector Quantization (LVQ) is used and at the second layer rough K-Means clustering algorithm has been used. The experimental results show that the effect is close to monolayer clustering algorithm than the rough K-Means, and the efficiency is better than the rough K-Means. A Hierarchical Representation Model with Multi-granularity (HRMM), which consists of five-layer representation of data and a two-phase clustering process, is proposed by Faliang *et al*.[11] based on granular computing and article structure theory. Roul *et al*.[12] in their work used Tf-Idf based apriori approach to cluster and rank the web documents. An effective clustering algorithm to boost up the accuracy of K-Means and spectral clustering algorithms is proposed in [13]. Their algorithm performs both bisection and merges steps based on normalized cuts of the similarity graph '*G*' to correctly cluster web documents. The proposed modified apriori approach scans the database only once and hence reduced the repeated database scan up to a maximum extent for making the initial clusters as compared to traditional apriori approach. The performance of the final clusters generated by FCM, VSM and K-Means techniques has been compared and the results show that FCM outruns VSM and K-Means.

## 3    Background

### 3.1    Traditional Apriori Algorithm

Apriori employs an iterative approach for finding frequent itemset. It starts by finding one frequent itemset and then progressively moves on to find (k+1)itemset from k-frequent itemset. From this resulting (k+1) itemset a set is formed which is the set of frequent(k+1) itemset. This process continues until no more frequent itemset are found. Apriori uses *breadth first search* algorithm for association. This can be used to group the itemset based on their association which satisfied the minimum support.

### 3.2.    Fuzzy C-Means Algorithm

FCM is a soft clustering technique which allows the feature vectors to belong to more than one cluster with different membership degree or belongingness of data. The value of the membership degree lies in the range from [0, 1]. It is simply an iterative fuzzy based algorithm that returns the centroid of the clusters. It will terminates when satisfy an objective function mentioned below.

$$T_m = \sum_{i=1}^{V} \sum_{j=1}^{C} u_{ij}^m \left\| d_i - c_j \right\|^2 \quad (1)$$

where,
m  [1, ∞] - fuzzy coefficient
$u_{ij}$ - membership degree of $x_j$ w.r.t to $c_j$; range [0, 1]
$c_j$ - centroid(vector) of cluster j

C - number of clusters
V - number of document vectors
$d_i$ - document vector

1. Initialize the membership matrix $U$ along with a fixed number of clusters. Number of rows and columns of $U$ depend on the number of clusters and documents. $U^{(0)} = [u_{ij}]$, where 0 is $0^{th}$ or first iteration.

2. Calculate the cluster center vector as $c_j = \frac{\sum_{i=1}^{V} u_{ij}^m \cdot d_i}{\sum_{i=1}^{V} u_{ij}^m}$  (2)

   where, all symbols have same meaning as defined in Equation 1. The membership values are calculated with respect to the new centers. Belongingness of the document to the cluster is calculated using Euclidian distance between the center and the document vector.

3. Update the $U^{(k)}$ matrix such that $u_{ij} = \dfrac{1}{\sum_{k=1}^{c} \left( \dfrac{\|d_i - c_j\|}{\|d_i - c_k\|} \right)^{2/(m-1)}}$  (3)

   where, all symbols having same meaning as in Equation 1 and 'k' is the iteration step.

4. If $\|U^{(k+1)} - U^{(k)}\| < $ , where,  $< 1$ is the termination criterion, then stop, else go to step 2. These will coverage to a local minimum of the function $T_m$.

### 3.3. K-Means Algorithm

1. Collect the documents $d_1, d_2, \ldots, d_k$ which need to be clustered.
2. Pre-determine the number of clusters ($K$).
3. Initialize the means of those pre-determined $K$ clusters.
4. Determine the Euclidean distance of each document from those means.
5. Group the documents having minimum distance to the corresponding mean.
6. Find the *new mean* of the each group formed in step 5.
7. If *new mean = previous mean* then stop else go to step 4.

### 3.4. Vector Space Model

Vector Space Model is an algebraic model which represents a set of documents as vectors in a common vector space. Thus, a document in its vector form can be look as, $D_i = [w_{1i}, w_{2i}, w_{3i}, \ldots\ldots\ldots\ldots, w_{ni}]$ where, $w_{ji}$ is the weight of term $j$ in document $i$.
If $x = \{x_1, x_2, \ldots, x_n\}$ and $y = \{y_1, y_2, \ldots, y_n\}$ are two documents in Euclidean n-space, then the distance between $x$ to $y$, or from $y$ to $x$ is:

*euclideanDistance(x, y)* $= \sqrt{\sum_{i=1}^{n}(x_i - y_i)^2}$  (4)

### 3.4.1. TF-IDF

TF the Term Frequency, which measures how often a term appears in a document. Tf($t$) = (Number of times the term $t$ appears in a document $D$) / ( Total number of terms in the document $D$).

IDF the Inverse Document Frequency, which measures how important a term is. If one combine Tf-Idf[14] then it can use to measure the relevance and importance of the term to a document. Idf($t$) = log(Total Number of documents) / (Number of documents in which the term $t$ appears). *Tf - Idf = Tf * Idf*.

### 3.4.2. Cosine_Similarity Measure

In Information Retrieval (IR) to measure the similarity between any two documents, a technique called Cosine_Similarity[14] has been used extensively. If *A* and *B* are two documents then cosine similarity between them can be represent as follows:

$$cosineSimilarity(A,B) = \frac{(A.B)}{(|A||B|)} \quad (5)$$

If $\Theta = 0^0$, then the two documents are similar. As $\Theta$ changes from $0^0$ to $90^0$, the similarity between the two documents decreases.

### 3.5  F-Measure

It measures the system performance by combining the precision and recall of the system[2]. Precision is the fraction of the retrieved documents that are relevant and recall is the fraction of the relevant documents that are retrieved.

$$\text{F-measure} = 2 * \frac{(precision * recall)}{(precision + recall)} \quad (6)$$

## 4  Proposed Approach

### 4.1. Optimization open Traditional Apriori

- ➢ Apriori generates frequent candidate itemset by generating all possible candidate itemset and then checking which itemset cross the minimum support count. Whereas our algorithm uses the following rule: If an itemset occurs ( k-1) times in the set of (k-1) frequent itemset then only it is considered for $k^{th}$ frequent candidate itemset(since only then it has a chance to come in a 'k' sized frequent itemset) [15]. So, our algorithm rejects more number of unwanted candidate itemset than normal apriori (which are generally not going to be frequent in the next iteration) before the counting step.
- ➢ Apriori counts the occurrence of an itemset even if it does not appear in any of the frequent candidates but our algorithm removes that itemset from the array by setting 0 across it so that less number of checks is made for that itemset.
➢ Apriori does not take into account the unnecessary computations made if the size of the transaction is lesser than the size of the candidate itemset being generated. So to improve upon this, we ignore the corresponding transaction by putting a null across it [16].

All of the above statements basically remove the information which is no longer required from the 2D-array created initially and this reduces the unnecessary comparisons in the subsequent steps.

### 4.2. Finding Initial Cluster Centers

The proposed modified apriori algorithm for finding frequent itemset has been used to find the number of clusters and initial cluster centers, which has used for all the three techniques (FCM, K-Means and VSM), The frequent itemset generated are of frequency greater than the minimum support supplied by the user. In the generated frequent itemset of documents, the keywords are taken to be the transactions and the documents are the itemset. In this way the frequent itemset generated are the ones which have particular set of keywords in common and hence are closely related. This helps by deciding the number of clusters and also the centers of these clusters which is simply the centroid of the respective frequent itemset.

Algorithm 1: Modified Apriori Approach

**Input:** The dataset (D) and minimum support (min_sup)
**Output:** The maximum frequent itemset

1. Read the dataset into a 2D array and store the information of the database in binary form in the array with transactions as rows and itemset as columns.
2. $k \leftarrow 1$.
3. Find frequent itemset, $L_k$ from $C_k$, the set of all candidate itemset.
4. Form $C_{k+1}$ from $L_k$.
5. Prune the frequent candidates by removing itemset from $C_k$ whose elements do not come at least k-1 times in $L_k$.
6. Modify the entry in the 2D-array in memory to be zero for the itemset which are not occurring in any of the candidates in $L_k$.
7. Check the Size of Transaction (ST) attribute and remove transaction from 2D-array where ST<=k.
8. $k \leftarrow k+1$.
9. Repeat 5-8 until $C_k$ is empty or transaction database is empty.

*Step 5* is called the frequent itemset generation step. *Step 6* is called as the candidate itemset generation step and *step 7-9* are prune steps. Details of first two steps are described below.

*Frequent itemset generation*: Scan D and count each itemset in $C_k$, if the count is greater than min_sup, then add that itemset to $L_k$.

*Candidate itemset generation:* For k = 1, $C_1$ = (all itemset of length = 1).
For k > 1, generate $C_k$ from $L_{k-1}$ as follows:
The join step:
$C_k$ = k-2 way join of $L_{k-1}$ with itself.
If both $\{a_1,...,a_{k-2}, a_{k-1}\}$ & $\{a_1,..., a_{k-2}, a_k\}$ are in $L_{k-1}$, then add $\{a_1,...,a_{k-2}, a_{k-1}, a_k\}$ to $C_k$.

---

Algorithm 2: Finding Initial Clusters

---

**Input:** Documents set to be clustered, the user query
**Output:** Initial Clusters ($C_i$)

1. Web page extraction and preprocessing*:* Submit the query to a search engine and extract top 'n' web pages. Remove the stop and unwanted words. Select *noun* as the keywords. Stemming the keywords and stored the pre-processed 'n' pages as document, $D_i$, where $i = 1, 2, ....., n$. Consider each keyword as a transaction and $D_i$ as transaction elements(itemset).
2. Creation of Term-Document matrix(T) and finding Tf-Idf of $D_i$:
   Term document matrix, *T*, is created by counting the number of occurrences of each term (i.e keyword) in each document $D_i$. Each row $t_i$ of *T* shows a term's occurrence in each document $D_i$. Finding out the Tf-Idf of $D_i$ from *T* ($D_{inew} \leftarrow$ Tf-Idf($D_i$)).
3. Extraction of maximum frequent itemset:
   Algorithm 1 is used to extract the maximum frequent itemset (i.e documents) from *T* and each maximum frequent itemset is treated as one cluster which gives the initial clusters($C_i$).

---

Algorithm 3: Final Clusters using FCM

---

**Input:** Initial clusters $C_i$, Value of fuzziness parameter *m*, document vector $D_{inew}$ from Algorithm 2
**Output:** Final clusters $FC_i$

1. Centroid calculation //find cluster centroids
   **for** each frequent set $f_i \in C_i$ **do**
       $c_i \leftarrow$ Centroid;
       $k \leftarrow$ length($f_i$);
       **for** each document $d_j \in f_i$ **do**
           $c_i \leftarrow c_i + d_j$;
       **end**
       $c_i \leftarrow c_i / k$
   **end**

2. Assignment documents to their respective clusters
   Final clusters $FC_i$ can be obtained by applying FCM algorithm (discussed in section 3.2) on the set of documents $D_{inew}$, using initial cluster $C_i$, the centroid $c_i$, and the fuzziness parameter *m*.

**Algorithm 4: Final Clusters using VSM**

---

**Input:** Initial clusters $C_i$, document vector $D_{inew}$ from Algorithm 2
**Output:** Final clusters $FC_i$

1.  Centroid calculation //find cluster centroids
    **for** each frequent set $f_i \in C_i$ **do**
        $c_i \leftarrow$ Centroid;
        $k \leftarrow$ length($f_i$);
        **for** each document $d_j \in f_i$ **do**
            $c_i \leftarrow c_i + d_j$;
        **end**
        $c_i \leftarrow c_i / k$
    **end**
2.  Assign the documents to their respective clusters
    **for** each $d_i \in D_{inew}$ **do**
        **for** each $c_j \in C_j$ ( i.e centroid) **do**
            Similarity[i][j] = *cosineSimilarity*($d_i$, $c_j$);
        **end**
        // find maximum similarity cluster($C_k$, k∈j) among the clusters from 0
        // to j-1.
        $C_k \leftarrow$ max(Similarity[i][0]….Similarity[i][j-1]); $C_k$.append($d_i$);
        //initial clusters get updated .
    **end**
3.  Repeat *step 1* and *2* until no cluster changes and it gives $FC_i$.

---

**Algorithm 5: Final Clusters using K-Means**

---

**Input:** Initial clusters $C_i$, document vector $D_{inew}$ from Algorithm 2
**Output:** Final clusters $FC_i$

1.  Centroid calculation //find cluster centroids
    **for** each frequent set $f_i \in C_i$ **do**
        $c_i \leftarrow$ Centroid;
        $k \leftarrow$ length($f_i$);
        **for** each document $d_j \in f_i$ **do**
            $c_i \leftarrow c_i + d_j$;
        **end**
        $c_i \leftarrow c_i / k$
    **end**
2.  Assign the documents to their respective clusters
    **for** each $d_i \in D_{inew}$ **do**

```
        for each c_j ∈ C_j ( i.e centroid) do
            Similarity[i][j] = euclideanDistance(d_i, c_j);
        end
        // find maximum similarity cluster(C_k, k∈j) among the clusters from 0
        // to j-1.
            C_k←max(Similarity[i][0]….Similarity[i][j-1]);C_k.append(d_i);
            //initial clusters get updated .
    end
3.  Repeat step1 and 2 until no cluster changes and it gives FC_i.
```

## 5  Experimental Results

For experimental purpose, to demonstrate the effectiveness of our approach, Classic3 and Classic4 dataset of Cornell University [17] have been used. The dataset contains four categories (CISI, CRAN, MED and CASM) set of documents. Each category set contains over 1000 documents, but for our experimental analysis, the traditional and modified apriori algorithm were run over CISI, CASM and MED document sets by using 200,400,600,800 and 1000 documents from each set. The graphs depicted in Fig. 1, Fig. 2 and Fig. 3 suggests that the proposed algorithm has a better running time than the traditional apriori algorithm. The support count used for CASM data set is 5, for CISI is 10 and for MED is 25. It is clear from Fig. 4 that even on varying the support count for the dataset (600 documents of MED) the modified algorithm still outperforms the tradition apriori algorithm. The modified apriori algorithm focuses on limitations of the traditional apriori and as can be seen from the graphs that it outruns the traditional apriori algorithm. It can also be observed from these graphs that as the number of documents increases the difference in running time between the traditional and modified apriori algorithm becomes significant. Finally FCM, K-Means and VSM techniques are applied on the initial clusters formed by the modified apriori approach on CISI document set and the F-measure has been calculated separately for clusters of different sizes shown in Fig. 5. The results show that the performance measurement of FCM is better than VSM and K-Means for 400, 600, 800 and 1000 documents of clusters.

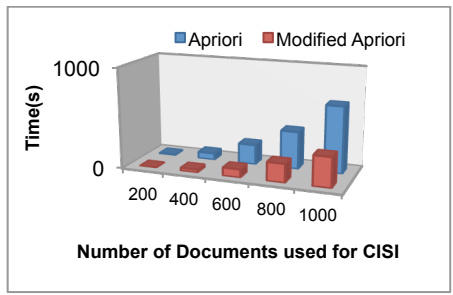
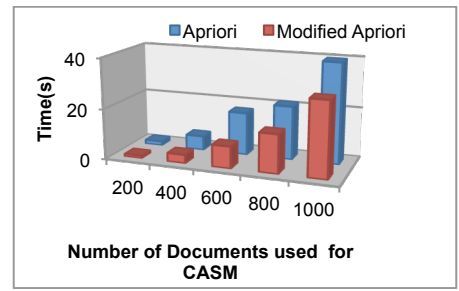

**Fig.1.** Apriori vs Modified Apriori(CISI)    **Fig.3.** Apriori vs Modified Apriori(CISI)

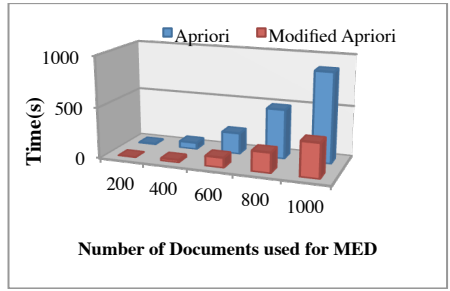
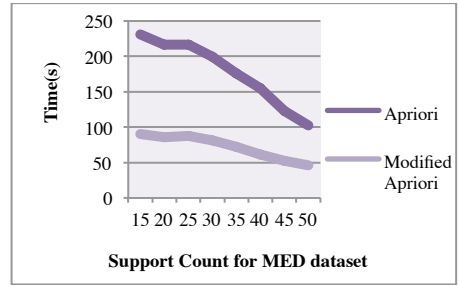

**Fig.2.** Apriori vs Modified Apriori(MED)     **Fig.4.** Support Count Graph(MED)

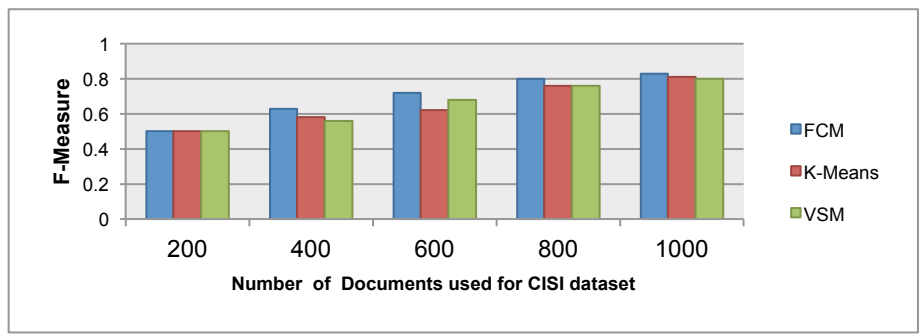

**Fig. 5.** Comparison of F-measure of different techniques(CISI)

## 6      Conclusion and Future Work

The proposed modified apriori approach when applied to a corpus of web documents produced the same clusters which is what a traditional apriori approach can produce. But theoretically and experimentally it has been shown that it is more efficient and faster than traditional apriori. This is so because at each step the information (*i.e* documents) from the corpus (*i.e* from the transactions) has been removed which is no longer required and in turn it reduces the unnecessary comparisons. Also the additional pruning of frequent itemset reduces the number of candidate itemset for the counting step and the number of database scans is reduced to *one*, since once the vertical data layout in the form of a 2D-array is made then no more database scans are required and all above manipulations are done on that 2D-array. Hence it saves a lot of time. After the initial clusters formed by the modified apriori approach, then FCM, K-Means and VSM techniques have been applied on it separately. Clasic3 and Classic4 dataset of Cornell University with document set of different size have been considered for experimental purpose. From the results it has been found out that the proposed approach is better than traditional apriori. Performance measurement in terms of F-measure has been carried out for final clusters produced by each of the three techniques. We found that FCM can able to give better clusters compared to K-Means and VSM. This work can be extended by labeling each cluster and ranking the documents for every cluster so that user can restrict their search to some top documents instead of wasting their time to search the entire cluster. Further text summarization can also be done on each cluster to get information about the content of that cluster.